\documentclass[10.75pt, a4paper]{article}
\usepackage{titlesec}
\titleformat{\section}
  {\bf\sffamily}
  {\thesection. }
  {5pt}
  {\MakeUppercase}
\renewcommand{\thesection}{\Roman{section}} 

\titleformat{\subsection}
  {\bf\sffamily}
  {\thesubsection. }
  {5pt}{}
  
\renewcommand{\thesubsection}{\Alph{subsection}}

\usepackage[affil-it]{authblk} 
\usepackage{etoolbox}
\usepackage{lmodern}

\usepackage{eurosym}

\usepackage[utf8]{inputenc}
\usepackage{geometry}
\usepackage[hidelinks]{hyperref}
\usepackage[citestyle=numeric,style=phys,biblabel=brackets,backend=bibtex,sorting=none,url=false,doi=false, natbib]{biblatex}
\bibliography{bibliography}
\DefineBibliographyStrings{english}{andothers={\itshape et\addabbrvspace al\adddot}}

\usepackage{csquotes}
\usepackage[]{authblk} 
\usepackage{etoolbox}
\usepackage{lmodern}

\usepackage{amsmath}
\usepackage{enumerate}
\usepackage{enumitem}
\usepackage{graphicx}
\usepackage{siunitx}
\usepackage{float}
\usepackage[]{parskip} %skip=0.7em, indent=1.5em

\makeatletter
\patchcmd{\@maketitle}{\LARGE \@title}{\fontsize{16}{19.2}\selectfont\@title}{}{}
\makeatother

\usepackage[normalem]{ulem}

\usepackage{graphicx}% Include figure files
\usepackage{dcolumn}% Align table columns on decimal point
\usepackage{bm}% bold math
\usepackage{tikz}
\usetikzlibrary{math}
\usetikzlibrary{shapes.geometric, arrows}
\usetikzlibrary{positioning}
\usetikzlibrary{shapes,arrows}
\usetikzlibrary{intersections}
\usepackage{csvsimple}
\usepackage{changepage}
\usepackage[tbtags]{mathtools}
\usepackage{siunitx}
\usepackage{csquotes}

\usepackage{float}
\usepackage{subfigure}
\usepackage[
	nonumberlist, 				% keine Seitenzahlen anzeigen
	acronym,      				% ein Abk�rzungsverzeichnis erstellen
	nomain,						% no main glossary
	nopostdot,					% Kein punkt nach Beschreibung
]{glossaries}
\usepackage{glossary-superragged}
\usepackage[hidelinks]{hyperref}
\usepackage{color, colortbl}

\usepackage{wrapfig}

\usepackage{pgfplots}
\usepackage{tikz}
\usetikzlibrary{arrows}
\usepackage{amsmath}
\pgfplotsset{compat=newest}
\usepgfplotslibrary{fillbetween}
\usetikzlibrary{calc}
\def\centerarc[#1](#2)(#3:#4:#5)% Syntax: [draw options] (center) (initial angle:final angle:radius)
{ \draw[#1] ($(#2)+({#5*cos(#3)},{#5*sin(#3)})$) arc (#3:#4:#5); }

\usepackage{amssymb}

\usepackage{nicefrac}

\usepackage{abstract}
\usepackage{multirow}
\usepackage{tabularx}
\usepackage{booktabs}
\usepackage{array}
\usepackage{longtable}
\newcolumntype{L}[1]{>{\raggedright\let\newline\\\arraybackslash\hspace{0pt}}m{#1}}
\newcolumntype{C}[1]{>{\centering\let\newline\\\arraybackslash\hspace{0pt}}m{#1}}
\newcolumntype{R}[1]{>{\raggedleft\let\newline\\\arraybackslash\hspace{0pt}}m{#1}}

\newacronym{3d}{3D}{three dimensional}
\newacronym{am}{AM}{additive manufacturing}
\newacronym{fdm}{FDM}{fused deposition modeling}
\newacronym{ism}{ISM}{in-space manufacturing}
\newacronym{iss}{ISS}{International Space Station}
\newacronym{fcb}{FCB}{Functional Cargo Block}
\newacronym{dem}{DEM}{discrete element method}
\newacronym{md}{MD}{molecular dynamics}
\newacronym{dc}{DC}{direct-current}
\newacronym[plural=PFCs,firstplural=parabolic flight campaigns (PFCs)]{pfc}{PFC}{Parabolic Flight Campaign}
\newacronym{fft}{FFT}{Fast Fourrier Transform}
\newacronym{cad}{CAD}{Computer Assisted Design}
\newacronym{ptfe}{PTFE}{polytetrafluoroethylene}
\newacronym{ps}{PS}{polystyrene}
\newacronym{nasa}{NASA}{National Aeronautics and Space Administration}
\newacronym{esamm}{ESAMM}{Extended Structure Additive Manufacturing Machine}
\newacronym{amf}{AMF}{Additive Manufacturing Facility}
\newacronym{us}{US}{United States}
\newacronym{usa}{USA}{United States of America}
\newacronym{bmgs}{BMGs}{Bulk Metallic Glasses}
\newacronym{esa}{ESA}{European Space Agency}
\newacronym{si}{SI}{International System of Units, abbreviated from French \textit{Syst\`{e}me International (d'unit\'{e}s)}}
\newacronym{dlr}{DLR}{German Aerospace Center}
\newacronym{liggghts}{LIGGGHTS}{\acrshort{lammps} Improved for General Granular and Granular Heat Transfer Simulations}
\newacronym{lammps}{LAMMPS}{Large-scale Atomic/Molecular Massively Parallel Simulator}
\newacronym{sjkr}{SJKR}{Simplified Johnson-Kendall-Roberts}
\newacronym{ded}{DED}{Directed Energy Deposition}
\newacronym{slm}{SLM}{Selective Laser Melting}
\newacronym{sls}{SLS}{Selective Laser Sintering}
\newacronym{eva}{EVA}{Extra-Vehicular Activity}
\newacronym{sem}{SEM}{Scanning Electron Microscopy}
\newacronym{RPM}{RPM}{Ramdom Positioning Machine}
\newacronym{rpm}{rpm}{revolutions per minute}
\newacronym{rise}{RISE}{Research Internships in Science and Engineering}
\newacronym{daad}{DAAD}{German Academic Exchange Service, abbreviated from German \textit{Deutscher Akademischer Austauschdienst}}
\newacronym{fsm}{FSM}{finite-state machine}
\newacronym{ir}{IR}{infrared}
\newacronym{pcbs}{PCBs}{Printed Circuit Boards}
\newacronym{pcb}{PCB}{Printed Circuit Board}
\newacronym{mcr}{MCR}{Modular Compact Rheometer}
\newacronym{sff}{SFF}{Solid Freeform Fabrication}
\newacronym{uv}{UV}{ultraviolet}
\newacronym{abs}{ABS}{acrylonitrile butadiene styrene}
\newacronym{hpde}{HPDE}{high density polyethylene}
\newacronym{pei}{PEI}{polyetherimide}
\newacronym{bff}{BFF}{BioFabrication Facility}
\newacronym{lens}{LENS}{Laser Engineered Net Shaping}
\newacronym{cnc}{CNC}{Computer Numerical Control}
\newacronym{ebf3}{EBF$^3$}{Electron Beam Free-Form Fabrication}
\newacronym{leo}{LEO}{Low Earth Orbit}
\newacronym{pc}{PC}{polycarbonate}
\newacronym{crissp}{CRISSP}{Customisable Recyclable International Space Station Packaging}
\newacronym{Athena}{Athena}{Advanced Telescope for High-ENergy Astrophysics}
\newacronym{lbm}{LBM}{Laser Beam Melting}
\newacronym{bam}{BAM}{Federal Institute for Materials Research and Testing, abbreviated from German \textit{Bundesanstalt f\"{u}r Materialforschung und-pr\"{u}fung}}
\newacronym{pbf}{PBF}{powder bed fusion}
\newacronym{eb}{EB}{Electron Beam}
\newacronym{2d}{2D}{two dimensional}
\newacronym{4d}{4D}{four dimensional}
\newacronym{ft4}{FT4}{Freeman Technology 4 Powder Rheometer}
\newacronym{dsc}{DSC}{Differential Scanning Calorimetry}
\newacronym{pmma}{PMMA}{polymethylmethacrylate}
\newacronym{1g}{$1g$}{gravity on-ground}
\newacronym{mug}{$\mu g$}{microgravity}
\newacronym{bcm}{BCM}{Box Counting Method}
\newacronym{mct}{MCT}{Mode Coupling Theory}
\newacronym{gmct}{gMCT}{granular Mode Coupling Theory}
\newacronym{itt}{ITT}{Integration Through Transients}
\newacronym{mfc}{MFC}{Mass Flow Controller}
\newacronym{ct}{CT}{computed tomography}
\newacronym{xct}{XCT}{X-ray computed tomography}
\newacronym{cv}{CV}{curriculum vitae}
\newacronym{pi}{PI}{principal investigator}
\newacronym{osp}{OSP}{orthogonal superimposed perturbation}
\newacronym{npi}{NPI}{Network Partnering Initiative}
\newacronym{ecsat}{ECSAT}{European Centre for Space Applications and Telecommunications}
\newacronym{eac}{EAC}{European Astronaut Centre}
\newacronym{estec}{ESTEC}{European Space Research and Technology Centre}
\newacronym{fps}{fps}{frames per second}
\newacronym{pdf}{pdf}{probability density function}
\newacronym{al}{Al}{aluminium}
\newacronym{ss}{\textit{SS}}{\textit{Smooth Surface}}
\newacronym{rs}{\textit{RS}}{\textit{Rough Surface}}
\newacronym{rcp}{rcp}{random close packing}
\newacronym{iop}{IoP UvA}{Institute of Physics of the University of Amsterdam}
\newacronym{mp}{MP}{Institute of Material Physics for Space}
\newacronym{elgra}{ELGRA}{European Low Gravity Research Association}
\newacronym{zarm}{ZARM}{Center of Applied Space Technology and Microgravity}
\newacronym{piv}{PIV}{particle image velocimetry}
\usepackage[framemethod=tikz]{mdframed}
\usepackage{lipsum}
\usepackage{dirtytalk}
\usepackage{tcolorbox}
\usepackage[font=footnotesize]{caption}
\newtcolorbox{mybox}[1]{colback=green!6!white,colframe=black!75!black,fonttitle=\bfseries,title=#1}
\newtcolorbox{mybox2}{colback=red!5!white,colframe=red!75!black}

\usepackage{pifont}

\usepackage{soul,xcolor}
\setstcolor{red}

\usepackage{xcolor,hyperref}
\hypersetup{
   colorlinks,
   linkcolor={blue!50!black},%{red!80!black},
   citecolor={blue!50!black},
   urlcolor={blue!80!black}
} 
\definecolor{mycolor}{rgb}{0.122, 0.435, 0.698}

\title{Optimal disk packing of chloroplasts in plant cells}
\author[1]{Nico Schramma\footnote{n.schramma@uva.nl, ORCID: 0000-0003-3887-3416}}
\author[2]{Eric R. Weeks\footnote{ORCID: 0000-0003-1503-3633}}
\author[1]{Maziyar Jalaal\footnote{m.jalaal@uva.nl, ORCID: 0000-0002-5654-8505}}

\affil[1]{Van der Waals-Zeeman Institute, Institute of Physics, University of Amsterdam, \protect\\
Science Park 904, Amsterdam, 1098XH, The Netherlands}
\affil[2]{Department of Physics, Emory University, Atlanta, Georgia 30322, USA}

\begin{document}
\definecolor{brickred}{rgb}{0.8, 0.25, 0.33}
\definecolor{darkorange}{rgb}{1.0, 0.55, 0.0}
\definecolor{persiangreen}{rgb}{0.0, 0.65, 0.58}
\definecolor{persianindigo}{rgb}{0.2, 0.07, 0.48}
\definecolor{cadet}{rgb}{0.33, 0.41, 0.47}
\definecolor{turquoisegreen}{rgb}{0.63, 0.84, 0.71}
\definecolor{sandybrown}{rgb}{0.96, 0.64, 0.38}
\definecolor{blueblue}{rgb}{0.0, 0.2, 0.6}
\definecolor{ballblue}{rgb}{0.13, 0.67, 0.8}
\definecolor{greengreen}{rgb}{0.0, 0.5, 0.0}
\begingroup
\sffamily
\date{}
\maketitle
\endgroup
\section*{Abstract}
%Mazi's version
Photosynthesis is vital for the survival of entire ecosystems on Earth. While light is fundamental to this
process, excessive exposure can be detrimental to plant cells. Chloroplasts, the photosynthetic organelles, actively move in response to light and self-organize within the cell to tune light absorption. 
%This presents significant challenges for the
%dynamic spatial self-organization of chloroplasts inside cells. 
These disk-shaped motile organelles
must balance dense packing for enhanced light absorption under dim conditions with spatial rearrangements to
avoid damage from excessive light exposure.
Here, we reveal that the packing characteristics of chloroplasts within plant cells show signatures of optimality.
Combining measurements of chloroplast densities and three-dimensional cell shape in the water plant
\textit{Elodea densa}, we construct an argument for optimal cell shape versus chloroplast size to achieve
two targets: dense packing into a two-dimensional monolayer for optimal absorption under dim light conditions and packing at the sidewalls for optimal light avoidance. We formalize these constraints using a model for random close packing matched with packing simulations of polydisperse hard disks confined within rectangular boxes. The optimal cell shape resulting from these models corresponds closely to that measured in the
box-like plant cells, highlighting the importance of particle packing in the light adaptation of plants.
Understanding the interplay between structure and function sheds light on how plants achieve efficient photo adaptation. It also highlights a broader principle: how cell shape relates to the optimization of packing finite and relatively small numbers of organelles under confinement. This universal challenge in biological systems shares fundamental features with the mechanics of confined granular media and the jamming transitions in dense active and passive systems across various scales and contexts.

\newpage

\section*{Introduction}

Photosynthesis is a fundamental process necessary for most life on Earth. However, fluctuations in light impose significant stress on plants, necessitating permanent dynamic adaptation. In addition to exhibiting macroscopic movements, such as phototropism, heliotropism, and shade avoidance~\cite{Forterre2013,Meroz2019,Moulton2020,Nguyen2024}, plants are also capable of changing their intracellular structure~\cite{Geitmann2015}, for example, by moving chloroplasts in response to light~\cite{Kasahara2002,Davis2011,Wada2018,schramma2023chloroplasts}.
These disk-shaped organelles, responsible for photosynthesis, can move towards or away from light by actively assembling networks of short actin filaments around them~\cite{Kadota2009,Suetsugu2010,Wada2018,Dwyer2022,Wada2024,kong2024}, allowing them to collectively rearrange the intracellular structure to tune the optical properties of plant cells~\cite{Senn1908,Zurzycki1961,Davis2011}.
Dim light eventually leads to an accumulation of chloroplasts in a layer to maximize the absorption of light~\cite{Gotoh2018}, while in strong light, chloroplasts move toward the sidewalls to increase leaf transmittance and avoid photodamage, such as increased production of reactive oxygen species~\cite{Park1996,Kasahara2002,Ruban2009,Li2009,Sztatelman2010,Davis2012}.
Although the molecular driving mechanisms of these movements are well studied, the collective aspects of the large-scale re-arrangement motion of chloroplasts in plant cells remain enigmatic: 
%How can a cell, constrained by a rigid cell wall and predominantly occupied by a central vacuole, reorganize its entire intracellular structure in response to light conditions, shifting seamlessly between dim and intense light environments?
How do cell shape and size impact the ability of the relatively large number ($N\approx50-120$) of chloroplasts to collectively re-arrange the intracellular structure to achieve various packing configurations for light adaptation?
%is scale movement of a large volume of chloroplasts between various packing configurations for dim and intense light environments.

Waterplants such as \textit{Elodea densa}, the subject of this study, provide an optimal system for microscopically studying chloroplast motion due to their simple two-cell-layered leaf structure. In a previous study, we identified a glass-like state under dim light conditions, where chloroplasts in a dense two-dimensional configuration (packing fraction $\phi \approx 70-74\,\%$) are caged and unable to move freely, exhibiting dynamics similar to those in glassy systems~\cite{schramma2023chloroplasts}. These mechanical characteristics stem from the high two-dimensional density of chloroplasts, which are bound to mostly move on the inner walls of the cells, as their movement relies on the anchorage to the plasma membrane~\cite{Whippo2011,Wada2018}, which can be alleviated via blue light in water plants~\cite{Ryu1995,Sakai2005,Sakai2015,Sakai2017}.
Upon strong light stimulation, the organelles become highly active and quickly transition within tens of minutes out of this two-dimensional glassy regime into a three-dimensional collective swirling motion of aggregates. These aggregates eventually spread on the side walls, enabling light avoidance (Supplementary Movie~1).

While the dynamic phases themselves pose intriguing questions about photo-activated phase transitions in a biological active matter system under confinement, several questions remain open regarding the underlying geometric aspects of chloroplast packing within cell confinement.
In fact, the dynamic adaptation response is infeasible if chloroplast number or size is altered. Normally, chloroplasts reach 2D packing densities of $\phi\approx54\,\%$ in spinach, $63\,\%$ in beetroot~\cite{Honda1971}, $69\,\%$ in wheat~\cite{Ellis1985} and around $80\,\%$ in Arabidopsis~\cite{Osteryoung1998}, with their number correlating to the area of their cells~\cite{Honda1971,Ellis1985} and chloroplast size~\cite{Pyke1992}.
% Light-dependent chloroplast motion is important to dynamically tune the optical properties of the leaf~\cite{Davis2011}.
% However, this adaptation response is infeasible if chloroplast number or size are altered: 
It was found that a few enlarged chloroplasts cannot re-arrange within the cell to reduce photo-damage efficiently~\cite{McCain1998,Koniger2008}, while a large population of smaller chloroplasts performs better~\cite{Jeong2002,Dutta2015,Dutta2017,Xiong2017}. These studies suggest that chloroplast size might be optimal for photosynthesis~\cite{Gowacka2023} and is crucial to be well controlled. Importantly, not only chloroplast number and size but also the cell shape have an impact on this adaptation response; for example, in lobed cells of \textit{Magnolia} or \textit{Zamia} leaves, chloroplasts cannot re-arrange efficiently~\cite{Davis2011}. These lead to the hypothesis that alterations to cell size and shape play a significant role in controlling the photo-protection efficiency. 

Here we study the interplay between cell geometry and chloroplast size to determine balanced packing to achieve optimal light harvest and photoavoidance motion in the cuboid cells of the waterplant \textit{Elodea densa}.
To accomplish this, we model chloroplast packing structures as a disk packing problem of two-dimensional polydisperse disks in rectangular confinement. The optimal packing of $n$-dimensional spheres ($n\geq2$) is a century-old problem with wide-ranging applications in condensed matter systems~\cite{Aste2008Book,Torquato2010,Torquato2010_RevMo}, optimization~\cite{Dowsland1992}, and signal transduction~\cite{shannon1948mathematical,sloane1984packing}.
The complexity of this seemingly simple problem is evident in the case of monodisperse spheres in three-dimensional free space. The optimal packing was conjectured to be a face-centered cubic structure by Johannes Kepler in 1611~\cite{Kepler1611} (but studied centuries earlier in a Sanskrit work ``ĀryabhatĪya of Āryabhata'' from 499 CE~\cite{Hales2006}) and was ultimately proven almost 400 years later by Thomas Hales in 1998~\cite{Hales2005,Aste2008Book}. Notably, packing (or tiling) constituents in living materials, similar to many classical condensed matter systems, feature additional complexities. First, in most cases,  the packing is disordered, lacking a clear crystalline structure or consisting of building blocks with more complex shapes~\cite{Classen2005,Gires2023,Fabreges2024,Ross2024,Skinner2023,Day2021,Srinivasan2023}.
%
% the packing of constituents is fundamentally vital in living materials. Many biological systems organize into complex packing (or tiling) patterns to perform various functions. Examples range over all domains of life include the packing structure of epithelial cell layers~\cite{Classen2005}, cytoplasmic structures~\cite{Gires2023}, cells in embryos~\cite{Fabreges2024}, avian photoreceptors on the retina~\cite{Jiao2014}, patterns on growing squid embryos~\cite{Ross2024},  bacterial colonies, zebrafish brain cells,the meristem of \textit{Arabidopsis}\cite{Skinner2023}, packing of yeast-cells and cells in the multicellular algae \textit{Volvox}~\cite{Day2021,Srinivasan2023}.
% While some of these systems exhibit order, many packings are disordered, lacking a clear crystalline structure or consisting of building blocks with more complex shapes. 
In the presence of noise and size-polydispersity, maximally dense packings can be determined algorithmically. However, the random close packing density $\phi_{rcp}$ is not well-defined and depends not only on particle shape~\cite{Roth2016,Yuan2018} and size distribution~\cite{Shimamoto2023,Anzivino2023,Meer2024} but also on the choice of the algorithm used~\cite{Torquato2000,Chaudhuri2010,Hermes2010}. Second, biological systems are often highly geometrically confined. Such a constraint strongly affects packing~\cite{Desmond2009,Edmond2012,Chen2015} and transport~\cite{Solano2024} .
By studying chloroplast configurations through the classical perspective of a packing problem, we aim to uncover critical dependencies between chloroplast size and cell shape. To achieve this, this work is divided into three parts: 
First, we quantify and describe the structure of cells in our model system. Next, we introduce the disk packing simulation based on experimental parameters. Finally, we construct two arguments for maximal packing in the cells' center and the side walls, which, when combined, define a clear optimal shape. 

\begin{figure}[t!]
    \centering
    \includegraphics[width=1\textwidth]{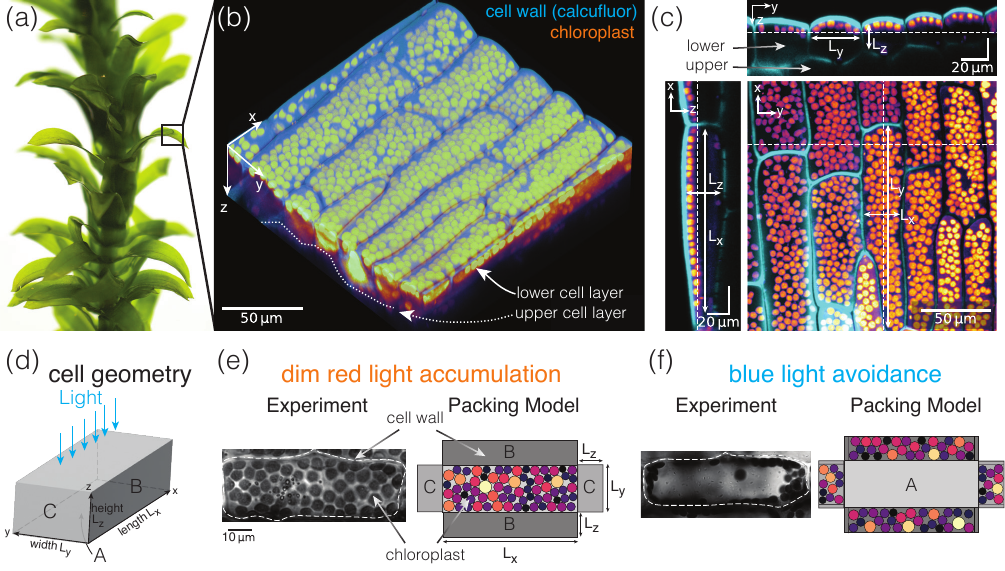}
    \caption{\textbf{Overview of experiment and packing model.} (a) The aquatic plant \textit{Elodea densa}. (b) A 3D confocal image of the cell walls shows a bi-layered leaf structure. We focus on the abaxial (lower) cell layer, depicted on top. Cyan channel: cell wall. Disk-like green structures: chloroplasts. Dotted line: guide to the eye to distinguish abaxial (lower) from adaxial (upper) cell layer. (c) Orthogonal view of cell walls (cyan) and chloroplasts (orange) along dotted white lines, respectively. Walls represent box-like structures with definitions of the box length $L_x$, width $L_y$, and height $L_y$ of the lower cell layer.
    (d) Schematic of the box geometry and experiment. Light enters from the top. Bottom wall A and sidewalls B and C are shown.
    (e) Cells under weak red light: chloroplasts are packed with a high density on the bottom wall. The packing model (right) shows a schematic of the opened box with chloroplasts packed solely in the A-plane. (f) Cells under strong blue light: chloroplasts spread on the side walls B and C. In the model, the chloroplasts are packed on B and C planes. A brighter color corresponds to larger chloroplasts.}
    \label{fig:Figure1}
\end{figure}

\section*{Disk-shaped chloroplasts are confined in elongated rectangular cells.}
\begin{figure}[t!]
	\centering
	\includegraphics[width=1.\textwidth]{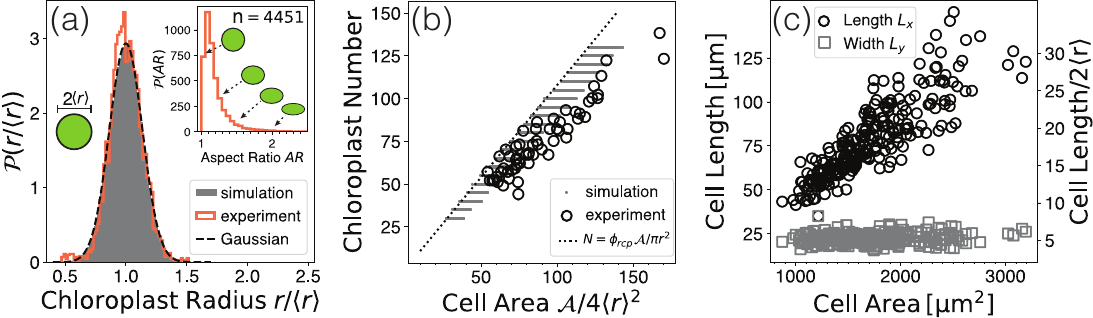}
	\caption{\textbf{Chloroplast and cell geometry}. (a) Gaussian statistics of chloroplast radii in experiments and simulations. Note: radii are normalized by the mean value, $\langle r \rangle$. Inset: the histogram of the aspect ratio shows that experimental chloroplasts are mostly circular. (b) Chloroplast number scales with cell area in simulation and experiment. Line depicts expected number for random close packing fraction $\phi_{rcp}=0.8478$ in free space. (c) Larger cells are more elongated (black circles) while having the same width (gray circles). Here, cell-dimensions are normalized with an average chloroplast diameter $2\langle  r\rangle = 4.25\,\mathrm{\mu m}$.}
	\label{fig:Fig2}
\end{figure}

We study the cell shape and structure of the water plant \textit{Elodea densa}, commonly referred to as ``waterpest'', using brightfield and confocal fluorescence microscopy. This monocot plant has a simple structure characterized by four-fold symmetric leaf arrangements on a single stalk (Fig.~\ref{fig:Figure1}a). The leaves have two layers: an adaxial (upper) layer with larger cells and an abaxial (lower) layer with cells approximately half the size (Fig.~\ref{fig:Figure1}b-c,  Fig.~\ref{fig:SI_Fig1})~\cite{Rascio1991,Hara2015} (similar to the related plant \textit{Elodea canadensis}~\cite{Ligrone2011}). 
We observe that upon strong light stimulation, the disk-shaped chloroplasts in the bottom of the cell move toward the cell walls after creating a motile aggregate (Fig.~\ref{fig:Figure1}d-f, Fig.~\ref{fig:SI_Fig2} and Supplementary Movie 1). 
% Furthermore, to elucidate the packing statistics in the dim light accumulated and strong light avoidant state, we perform disk packing simulations in confinement (\textit{Materials and Methods}) to mirror the packings in the face and the sidewalls, which we will come back to later.\\
To quantify the cell shapes and chloroplast sizes, we perform brightfield imaging and chlorophyll auto-fluorescence microscopy (\textit{Materials and methods}, Fig.~\ref{fig:SI_Fig3}). 
We analyze $262$ cuboid-shaped cells and $n=4451$ disk-shaped chloroplasts (in $59$ cells), with an approximately Gaussian-distributed disk-radius $r = 2.12 \pm 0.29 \,\mathrm{\mu m}$ (mean $\pm$ standard deviation) and polydispersity $\delta = \langle r\rangle/ \sigma_r = 13.6 \%$ and with an aspect ratio close to $1$ (Fig.~\ref{fig:Fig2}a). 
We find scaling between cell area $\mathcal{A}$ and chloroplast number $N$ (Fig.~\ref{fig:Fig2}b), consistent with previous observations in other plants~\cite{Honda1971,Ellis1985,Pyke1992,Gowacka2023}. As chloroplasts in dim light mostly pack in a single layer, the upper limit of this scaling is expected to result from random close packing in two-dimensional free space with a packing density $\phi_{rcp}$ (dotted line in Fig.~\ref{fig:Fig2}b).

However, the chloroplast number lies well below this line, which likely results from the dependency of random close packing on the confinement \cite{Desmond2009}.
Furthmore, we find, that cells have various lengths $L_x = 50 - 125\,\mathrm{\mu m}$ while their width remains largely constant $L_y = 22.2 \pm 2.95 \,\mathrm{\mu m}$ (Fig.~\ref{fig:Fig2}c).
To provide a physical intuition about the cell confinement, we re-scale all dimensions by the average chloroplast diameter $2\langle r\rangle$. This renders all length scales in terms of the average number of chloroplasts that fit within a given space and suggests that only $4-7$ chloroplasts fit within the width of the cells, while the cell length varies between approximately $10$ to $30$ chloroplasts.

To measure the height of the cuboid cells, we stain the cell walls with calcofluor and perform confocal microscopy to generate three-dimensional volumetric images of the plant cells  (\textit{Materials and Methods}). We find that the cells have an average height $L_z/2\langle r \rangle \approx 2.34\pm1$ ($n=86$) measured within one average chloroplast diameter from the boundary (compare Fig.~\ref{fig:Figure1}c, Fig.~\ref{fig:SI_Fig2}c,d).
Notably, sometimes the lower cell walls align with the upper cells, creating trapezoidal shapes in which one cell wall is much higher than the other (Fig.~\ref{fig:SI_Fig2}).
Taken together, this shows that chloroplasts are highly confined in two directions and have greater freedom to arrange along the long side of the cell. With these insights about chloroplast size and the respective confinement, we ask whether these sizes are optimally related to achieving both objectives: optimal light capture and optimal chloroplast avoidance. To address this, we will present a theoretical argument in the next section.

\section*{2D Chloroplast Packing in Confinement}
It has been observed that chloroplasts move toward the bottom wall under dim light and toward the sidewalls under strong light~\cite{Jeong2002,Sakai2005,Nagai1993,Wada2018} (see Fig.~\ref{fig:Figure1}e,f), with each configuration - chloroplasts in a single bottom layer or at the sidewalls - serving distinct purposes for light adaptation. The first is associated with the optimal light uptake under dim conditions~\cite{Gotoh2018}, while the latter is a light-avoidance response that optimizes the intracellular structure for enhanced light transmission and thus reduced photo-damage~\cite{Zurzycki1961,Kasahara2002,Davis2011}.
To outline the interplay of chloroplast number, size, and cell geometry, let us consider a cuboid cell (container) shaped to allocate many disk-shaped chloroplasts in the bottom layer (Fig.~\ref{fig:Figure1}d-f). If the side walls are not large enough to accommodate all chloroplasts during strong light avoidance, excessive light exposure may harm the chloroplasts. This occurs if the cuboid is relatively flat and the bottom and top faces being large squares. On the contrary, if the side walls are much larger compared to the bottom area (with the top and bottom faces being highly elongated rectangles), the chloroplasts would easily fit into the bottom layer, leaving significant empty space due to inefficient packing and boundary defects, which is suboptimal for the metabolite-production via photosynthesis.

Furthermore, we must consider that the disk-like organelles have an upper bound on their maximal packing density (random close packing), which depends on the confinement, similar to the packing of disks in a plane or spheres in a box~\cite{Desmond2009,Chen2015}. 
Here, we formalize these mathematical upper bounds for (I) random close packing and (II) the side-to-bottom area mismatch to find the optimal geometry for packing under both constraints.
We anticipate that experimental data will fall well below this upper bound, as the disks (chloroplasts) must dynamically re-arrange between configurations and, therefore, cannot be strongly jammed.

\subsection*{Constraint (I): random close packing in confinement}
\begin{figure}[t!]
	\centering
	\includegraphics[width=1\textwidth]{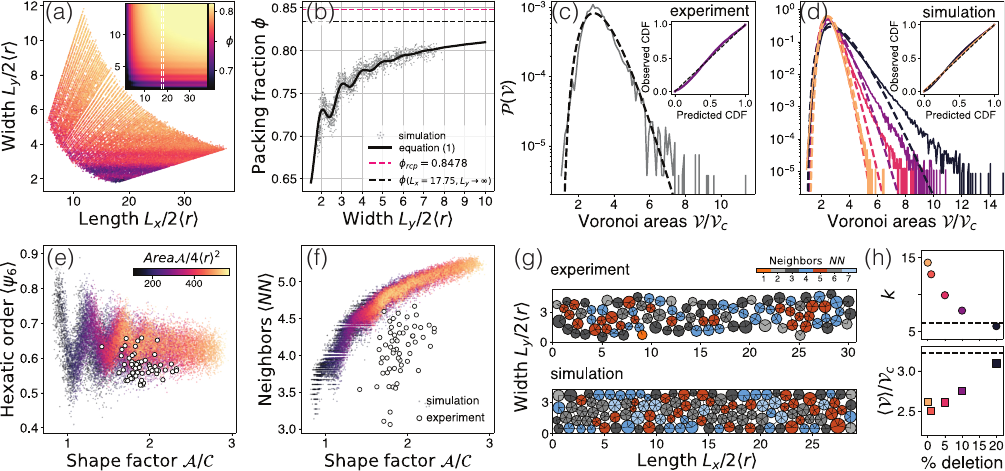}
	\caption{\textbf{Packing statistics of simulation and experiment} (a) Map of simulated maximal disk packing fraction in respective confinements ($L_x,L_y$). Inset: fit of equation (1). (b) Packing fraction of simulation (points) and fit (black solid line) depending on the width $L_y$, at cell lengths $L_x \in {17,18.5}$ (dashed white lines in the inset of (a)). Black dashed line: expected asymptotic packing fraction for uniaxial confinement $L_y\to\infty$. Magenta dashed line: expected asymptotic packing fraction in free space. Oscillations are significant at confinements of $L_y/2\langle r \rangle \lesssim 4$. (c) Voronoi area distribution of experiment packing follows a k-Gamma distribution (dashed line). Inset: P-P plot of cumulative distribution function (CDF) suggests slight variations of observed and maximum entropy distribution, suggesting structure. (d) Distribution of simulated packings with increasing random deletions of particles (darker colors), indicating that random voids can generate structure (see (h) for parameters).
    (e) Average hexatic order parameter for all confinements, indicated by a shape factor comparing area to perimeter. Colormap represents the area. Strong oscillations become apparent in smaller and more elongated cells. White-points: experimental observation. (f) The number of nearest neighbors scales with the size and shape of the container. The larger the bulk phase, the more neighbors are allowed. The number of nearest neighbors in experiments significantly underestimates that of the random packings, suggesting more defects and less dense packing.
    (g) Packing in experiments (top) and simulations (bottom), colors indicate the nearest neighbor numbers $NN$. Lines: nearest neighbor network. 
    (h) Distribution parameters for shape $k$ and average $\langle \mathcal{V}\rangle$ compared to experiment (dashed line) as a function of the percentage of random deletions. The colors of each point correspond to the data in (d).}
	\label{fig:Fig3}
\end{figure}

The packing fraction for a disordered arrangement of disks with radii $r$ drawn from a distribution $\mathcal{P}(r)$ in confinement can be approximated by the random close packing (rcp) fraction $\phi_{rcp}$. Although random close packing is not precisely defined and varies depending on the algorithm used~\cite{Torquato2000,Torquato2010_RevMo,Atkinson2014}, we employ this concept to estimate changes in packing density under confinement.
Our chloroplast data indicates a remarkably Gaussian distribution of radii with a polydispersity of $\delta  = \sigma_r/ \langle r \rangle = 13.6\,\% $ (Fig.~\ref{fig:Fig2}a) and a low average aspect ratio, $AR < 1.1$.
For a unimodal Gaussian distribution of diameters with such a polydispersity, the random close packing density in asymptotically free space is $\phi_{rcp} \approx 0.841$ ~\cite{Meer2024}. The value of the random close packing fraction for confined disks, nonetheless, is less trivial.
We simulate the disk packing in two-sided confinement, $L_x$ and $L_y$, using an approach adopted from~\cite{Meer2024} (\textit{Materials and Methods}), which is originally based on the work of Xu \emph{et al.}~\cite{Xu2005} and Clarke and Wiley~\cite{Clarke1987}, and which has also been applied to one-sided confinement~\cite{Desmond2009}. Running $23479$ simulations with varying confinement widths, $L_x  = 5.25-37.41$ and $L_y = 1.79-11.78$, we find that the random close packing fraction, $\phi$, depends on both the x- and y-confinement (Fig.~\ref{fig:Fig3}a).
We refer to the length scales here as $L_x$ and $L_y$, expressed in units of the average chloroplast diameter, $2\langle r \rangle$. To explore the variation in random close packing fraction $\phi$ within a 2D-container under confinement, we use a phenomenological relationship, first introduced in 1946~\cite{Verman1946,Brown1946}, and has also been applied to 1D confinement~\cite{Desmond2009} and rods in a cylindrical container~\cite{Freeman2019}: $\phi = \phi_{rcp}-\alpha(\frac{1}{L_x}+\frac{1}{L_y}) = \phi_{rcp}-\frac{\alpha}{2}\frac{\mathcal{C}}{\mathcal{A}}$.
Here, $\phi_{rcp}$ is the free-space random close packing fraction of the polydisperse disks, $\mathcal{C}=2(L_x+L_y)$ is the perimeter, and $\mathcal{A}=L_xL_y$ is the box area.
However, we observe significant oscillations in packing density under strong confinements, $L_y \lesssim 4$ (Fig.~\ref{fig:Fig3}a,b). These oscillations arise from integer mismatches of fitting the unimodally distributed disks, with maxima aligning at integer-values of $L_y = (1,2,...)$ and minima at  $L_y = (3/2,5/2,...)$ (Fig.~\ref{fig:Fig3}b).
To precisely model slender cells and especially the packing on the side walls, we extend the simple hyperbolic law by introducing a damped oscillatory correction term (Fig.~\ref{fig:Fig3}a, inset):

 \begin{align}
    \label{eq:packing_geometry}
     \phi_{I}(L_x,L_y) = \phi_{rcp}-\alpha\left(\frac{1}{L_x}+\frac{1}{L_y}\right) + \beta\left(\cos(2\pi L_x)e^{-L_x/\xi} + \cos(2\pi L_y)e^{-L_y/\xi} \right) 
\end{align}

The oscillatory relation is mainly needed to accurately model the packing on the side walls (x-z and x-y planes in Fig.~\ref{fig:Figure1}), where one dimension is highly confined ($L_z\approx2.34$). 
Fitting this symmetric relation to our simulation data gives $\phi_{rcp} = 0.8478$ for the free-space random close packing of this disk size distribution, with $\alpha = 0.2444$, $\beta = 0.0825$ and $\xi = 1.284$ (errors below $10^{-4}$).
This constraint serves as a mathematical upper bound for two-dimensional chloroplast packing within the cell. For confinements where $L_i<1$ ($i\in\{x,y\}$), the relation breaks down as sampling from a Gaussian with an average disk diameter of $2\langle r\rangle=1$ becomes strongly constrained (bigger disks cannot fit within the box). This reduces the effective average sampled radius, which is instead drawn from a truncated Gaussian distribution: $\langle r \rangle_L = \langle r \rangle - \sigma_rg(x)/\mathcal{G}(x) \approx 0.445$,  where $g(x) =\exp(-\frac{x}{2})/\sqrt{2\pi}$, $\mathcal{G}(x) = (1+\mathrm{erf}(x/\sqrt{2}))/2$, and $x=(L/2-\langle r \rangle)/\sigma_r$. 
In our case, this results in a packing fraction for such confinement at $L_y=1$: $\phi = \pi \langle r\rangle_L^2 \approx 0.622$. This explains the discrepancy of the fitted function (eq.~\eqref{eq:packing_geometry} which reaches $\phi_{I}(1,1) = \phi_{rcp} - 2\alpha -2\beta e^{-1/\xi}\approx 0.435$, largely underestimating the sampling-corrected packing fraction. Hence, the estimation of the mean needs to be corrected for very strong confinements, though this correction is not considered here as all length-scales remain $L_i > 1$.

\subsection*{Constraint (II): area-side-wall mismatch}

The ability of chloroplasts to move towards the side walls under strong light introduces a second geometrical constraint on the maximal possible packing fraction.
If the chloroplasts can cover the area $\mathcal{A}=L_xL_y$ at a high packing fraction $\phi_A(L_x,L_y)$, they must also be able to cover the four sidewalls without exceeding the maximal packing fractions $\phi_{B}$ and $\phi_{C}$ (see Fig.~\ref{fig:Figure1}g,h).
We obtain the theoretical maximal packing density at the walls, $\phi_{B}, \phi_{C}$, approaching the respective random close packing $\phi_{B}\to\phi_{I}(L_x-1,L_z)$ and $\phi_{C}\to\phi_{I}(L_y-1,L_z)$, as described in eq.~\eqref{eq:packing_geometry}. Note that at the side walls, the effective wall length and width are reduced by one chloroplast diameter as the chloroplasts are positioned along the inner walls. If the sidewalls were fully packed and folded into a box, chloroplasts might overlap due to their three-dimensional shape (see Fig.\ref{fig:Figure1}d-f for comparison).\\

Henceforth, we require that the maximal number of chloroplasts covering the bottom area also fits onto the side walls $N \pi/4 =\phi_A \mathcal{A} \leq 2L_z\left(\phi_{B}\cdot(L_x-1)+\phi_{C}\cdot(L_y-1)\right)=N_{side}\pi/4$. Here, the area of the chloroplast $\pi \langle r \rangle ^2$ has been normalized by the area of a square containing it $4\langle r \rangle^2$. Using equation \eqref{eq:packing_geometry} we arrive at:
\begin{align}
\label{eq:packing_sidewalls}
    \phi &\leq \phi_{II}=\frac{2L_z}{\mathcal{A}}\left(\phi_I(L_x-1,L_z)(L_x-1)+\phi_I(L_y-1,L_z)(L_y-1)\right) .
\end{align}

The density $\phi$ at the bottom is subject to two constraints: (I) the geometrically feasible random close packing (eq. \eqref{eq:packing_geometry}) and (II) the available space at the side walls (eq. \eqref{eq:packing_sidewalls}). Consequently, the maximal possible packing under both constraints $\phi$ must satisfy:
\begin{align}
    \label{eq:phimax}
    \phi \leq \min(\phi_{I},\phi_{II}) \equiv \phi^*. %
\end{align}

\subsection*{Structural comparison of chloroplast packing and simulations}
We compare simulations and the model $\phi^*$ top experiments by analyzing the data from $59$ cells containing $4451$ chloroplasts to determine whether their structural properties are similar to those of the simulated packing configurations. To this end, we perform a Voronoi-tesselation of the chloroplast positions in confinement (Fig.~\ref{fig:Fig3}c,d). The distribution of Voronoi areas $\mathcal{V}$ shows a slight deviation from the k-Gamma distribution, with a shape parameter $k$, a cutoff scale $\mathcal{V}_c$ and an average $\langle \mathcal{V} \rangle$. The comparison with the k-Gamma distribution is primarily done since it represents a maximum entropy law found in various packing structures of granular media and cells~\cite{Aste2008,Day2021}. While the k-Gamma law closely fits the simulated packing structures (by comparing cumulative distribution function (CDF) of data and the k-Gamma distribution, see inset of Fig.~\ref{fig:Fig3}d), introducing a small number of random deletions of disks($1\,\%$ to $20\,\%$ of the disk number) produces deviations in the probability distribution function similar to that in the experiments (Fig.~\ref{fig:Fig3}c,d,h). This suggests that the measured chloroplast packing is more representative of packing densities below random close packing.\\

Further, we quantify the $p$-atic bond-orientational order parameter ($p\in\{4,5,6,7\}$) for each chloroplast $j$ and its $N_j$ nearest neighbors, then average the results within each cell 
\begin{align}
\label{eq:hex}
    \langle \Psi_p \rangle = \langle \frac{1}{N_j}\sum_{k=1}^{N_j} e^{ip\theta_{jk}} \rangle,
\end{align} 
where $\theta_{jk}$ is the bond-angle between particle $j$ and its nearest neighbor $k$. Focusing on the hexatic order parameter ($p=6$) and the number of nearest neighbors (Fig.~\ref{fig:Fig3}e-g), we find that, although the nearest neighbor number is slightly lower than in the simulations, the hexatic order is similar. This suggests a comparable overall structure: a disordered material with a few hexagonal domains (compare also $p=4,5,7$ in Fig.~\ref{fig:SI_Fig4}). Strikingly, these values are consistent with those reported in~\cite{Meer2024}, where polydispersity was shown to have a strong effect on hexatic order.
Additionally, we find that most packings exhibit an average nearest-neighbor number of $N>4$ (similar to a coordination number), which is a feature of mechanical stability of jammed structures in two dimensions~\cite{Mason1960,Torquato2010_RevMo}. Next, we use the cell shape data to compare it to the predictions from the model $\phi^*$.

\subsection*{Cell shape is optimal for chloroplast packing}
\begin{figure}[h!]
\centering
\includegraphics[width=1\textwidth]{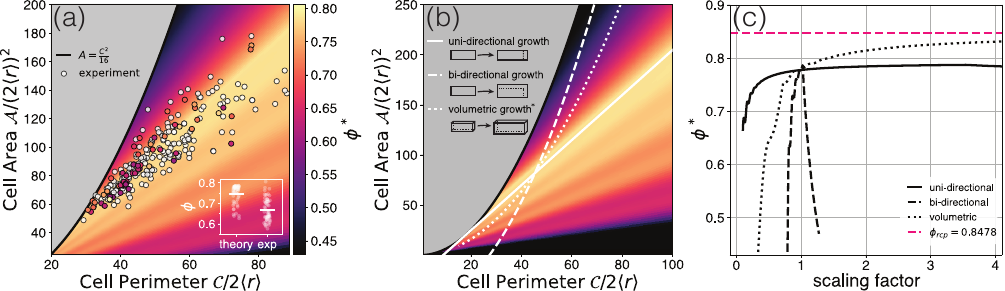}
	\caption{\textbf{Maximal packing under geometric constraints.} (a) Color represents the maximal possible packing fraction $\phi^*$ according to eq. \eqref{eq:phimax}. The shaded area is inaccessible to rectangles and exceeds the formula for squares $A=C^2/16$. Circles represent individual cells from microscopy experiments, with packing fractions indicated by color. Inset: comparison of theoretically expected maximal packing fraction and measured packing fraction for all cells. (b) Same field as in (a) with extended boundaries. Lines indicate different growth models: uniaxial (solid line, $L_x$), bi-axial (dashed line, $L_x$,$L_y$), and volumetric growth models (dotted line, $L_x$,$L_y$,$L_z$). Volumetric up or down-scaling of the cell size is equivalent to down or up-scaling of chloroplast size, respectively. (c) Measured maximal packing fraction for the three growth models with corresponding lines. The scaling factor compares the updated to the original cell length: $L_x'/L_x$. Dashed magenta line: random close packing fraction in free space.}
	\label{fig:Figure_double_constraint}
\end{figure}

We represent the coordinates of the field $\phi^*(L_x,L_y,L_z)$ in terms of the dimensionless cell area $\mathcal{A}$ and perimeter $\mathcal{C}$ (Fig.~\ref{fig:Figure_double_constraint}a), while keeping $L_z=2.34$ constant. This representation comes with the caveat of an excluded region, as $\mathcal{A}\leq \mathcal{C}^2/16$ for rectangles (the maximal area for a given perimeter corresponds to a square).
%
% The maxima of $\phi^*$ are organized along a specific line for which $\phi_I = \phi_{II}$ if $\beta=0$. 
% We can represent this line as the transcendental equation:
%$\Leftrightarrow A=\left(L_z-\frac{\alpha}{2\phi_{rcp}}\right)C-\frac{4\alpha L_z}{\phi_{rcp}}$. 
Intuitively, constraint (I) is weaker for larger cells, as a large bulk phase allows for more possibilities of optimized packing (Fig.~\ref{fig:SI_Fig5}a). 
We can see that $\phi_I\to\phi_{rcp}$ as $L_x,\,L_y\to\infty$, i.e. for infinitely large boxes.
In contrast, constraint (II) is weaker for elongated and small cells, which have a larger sidewall area relative to the bulk area (Fig.~\ref{fig:SI_Fig5}b). This is evident by extending only one dimension, $L_x\to\infty$ while $L_y\to 0$, where $\phi_{II} \to \infty$, rendering this constraint irrelevant for very slim and elongated cells. On the contrary, $\phi_{II}\to 0 $ when both $L_x,\,L_y\to\infty$, i.e. a very strong constraint.
As both constraints are incompatible, we find a maximum ridge of optimality, where the packing fraction is as large as $81\%$, which is only $3.6\%$ below rcp in free space (Fig.~\ref{fig:Figure_double_constraint}a).

Intriguingly, the data of cell shapes coincides closely with this maximum ridge, suggesting that cells are compatible with optimal packing density. The measured chloroplast packing fractions (colored points in Fig.~\ref{fig:Figure_double_constraint} and inset) remain well below the expected maximal packing fraction. 
In fact, the average packing fraction of approximately $67\%\pm6\%$ (mean $\pm$ standard deviation) is around $10\%$ below the maximal packing fraction, suggesting that the cells could indeed occupy a larger space in the $\mathcal{C}-\mathcal{A}$-plane without losing much of the geometric benefits, or alternatively, accommodate a higher number of chloroplasts.

This discrepancy, however, is expected and can be explained by the difference between the average nearest neighbor distance $d$ and the chloroplast diameter $\langle d - 2r\rangle/2\langle r \rangle = 0.1 \pm 0.05$. Hence, there is a small inter-chloroplast distance of $l=0.42\pm 0.2\,\mathrm{\mu m}$, consistent with the previous observations~\cite{schramma2023chloroplasts}. This suggests that chloroplast packing is not organized directly at the close packing point but slightly below it, allowing for occasional re-arrangements of the chloroplasts over larger timescales~\cite{schramma2023chloroplasts}.
Additionally, other organelles and structures, invisible to our imaging method, need space, making full chloroplast contact highly unlikely.

Moreover, we found that some cells exhibit a more irregular height profile, especially when one cell wall of the lower cell layer is aligned with one of the upper cell layers. In such cases, the cell wall can form a deep trench of up to $L_z \approx 4$. Here, the anticlinal wall (facing outward, see Fig.~\ref{fig:Figure1}d A-side) and the periclinal walls (between the cells Fig.~\ref{fig:Figure1}d and B-side) have approximately the same area and shape. This configuration minimizes the influence of the side wall area as a constraint on the packing fraction.

In the last step, we illustrate lines of growth within the three-variate function $\phi^*(L_x,L_y,L_z)$. 
We model growth by starting at a specific cell size, close to optimality with $L_x = 18.2$, $L_y=4.5$ and $L_z=2.34$, and linearly increasing the size of the axes as $L_i\to L_i+\alpha_i \;,i\in\{x,y\}$. We model uni- and bi-directional growth by setting $\alpha_i$ to either $0$ or a linear growth function in time, $\alpha(t)$, for the $i$'th direction, respectively. This generates various curves in the $\mathcal{C}-\mathcal{A}$ plane (Fig.~\ref{fig:Figure_double_constraint}b).
Additionally, we scale all axes proportionally in all directions ($L_i\to \alpha_i L_i,\; \forall i \in \{x,y,z\}$), equivalent to changing the chloroplast size, as studied experimentally~\cite{McCain1998,Koniger2008,Jeong2002,Dutta2015,Dutta2017,Xiong2017,Gowacka2023}. 

We monitor how $\phi^*$ evolves on these curves, noting that volumetric scaling evolves in three dimensions. The corresponding $\phi^*$ function, dependent on the scaling factor (updated length compared to initial length $L_x'/L_x$), is shown in Fig.~\ref{fig:Figure_double_constraint}c. While uni-directional growth allows cells to remain aligned with the maximal ridge of $\phi^*$, bi-directional growth quickly deviates from the maximum, resulting in effectively more quadratic cell shapes. Notably, most cells have similar widths but largely different lengths (Fig.~\ref{fig:Fig2}c), suggesting uni-directional cell growth, enabling cells to remain near the maximal ridge during development.

Interestingly, volumetric scaling, which is equivalent to changing the size of chloroplasts, can even increase the maximal random closed packing fraction $\phi^*$. This could explain the observed adaptability enhancement in cells with smaller, more numerous chloroplasts~\cite{Jeong2002,Xiong2017}, as opposed to intercellular ``crowding''~\cite{McCain1998}. 
Our analysis, therefore, provides a potential framework to understand chloroplast motion and adaption under confinement and in relation to cell geometry.

\section*{Discussion}

Packing objects into a confined space is a challenging and ubiquitous problem~\cite{Aste2008Book}, from candies in a jar to understanding structural configurations in condensed matter systems~\cite{Torquato2010_RevMo,Royall2023}. The problem also extends to practical considerations such as packaging optimization~\cite{Dowsland1992} and more abstract scenarios like high-dimensional sphere packing~\cite{shannon1948mathematical,viazovska2017sphere}, which plays a role in telecommunications and information theory. Notably, biological systems have evolved to tackle packing challenges, from densely packed cells in embryos~\cite{Fabreges2024}, organoids~\cite{Tan2024} and tissues~\cite{Classen2005,farhadifar2007influence,tang2024tunable}, bacterial growth in confinement~\cite{Sreepadmanabh2024}, to compactly folded genetic material in nuclei~\cite{webster2009sizing,dougan2018three}.

We show that the packing of mesoscale structures, such as organelles in biological systems, plays a crucial role in adaptation processes. 
Specifically, we combined structural analysis of photosynthetic chloroplasts in the water plant \textit{Elodea densa} with disk-packing simulations in confinement to explore the efficiency of chloroplast packing within cells. To cope with the everchanging light conditions, chloroplast rearrangement results in two distinct configurations: high coverage on the bottom cell wall for maximal light absorption in dim conditions and a fast and efficient relocation to the side walls to minimize light exposure under strong light.

First, we investigated the size and shape of cells and chloroplasts. Notably, we found that chloroplasts are Gaussian distributed, which must rely on a size control mechanism. Simple growth-division models with either size-additive noise or size-multiplicative noise predict Gaussian and log-normal size distributions, respectively~\cite{Amir2014,Ho2018}. However, due to the small coefficient of variation (polydispersity $\delta=13.6\%$) these two distributions are not distinguishable from each other within the resolution of our data. 
Additionally, we found that the chloroplast number scales with cell size, consistent with observations in other plants~\cite{Honda1971,Ellis1985,Pyke1992}. This suggests an intrinsic mechanism that regulates chloroplast number and thus density~\cite{Pyke1999,Cackett2022}.
Furthermore, the cuboid cells display variations in lengths but similar widths and heights, a pattern consistent with unidirectional growth, as expected from development in monocot plants with simple leaf architecture, where growth is predominantly localized near the leaf base~\cite{Mathur2004,Nelissen2016}. 

To investigate the optimality of cellular geometry for chloroplast packing, we simulated the random close packing of disks under confinement. The geometry-dependent maximal packing was then compared to a phenomenological packing model, similar to the hyperbolic laws previously applied~\cite{Brown1946,Verman1946,Desmond2009,Freeman2019}, with an added explicit treatment of non-monotonic deviations under strong confinements ($L\lesssim 4$). This was necessary for accurately evaluating packing on the side walls with heights around $L_z\approx 2.34$.
By combining two constraints - maximal packing on the bottom side and all side walls of the box, we constructed a law for cell shape-dependent optimal packing. Mapping measured cell structures to this morphological packing criterion showed that the cuboid cells have optimal shape and dimension to meet these two targets. This morphological feature allows cells to adapt their intracellular structure efficiently for optimized light absorption while simultaneously being able to mitigate potential photo-damage by switching between packing configurations. We hypothesize that this trait is likely a result of evolutionary adaptation to the plant's highly fluctuating aquatic environment. Furthermore, this simple physical constraint may also explain the inefficient chloroplast re-arrangement observed in various mutant plants with altered chloroplast sizes and numbers~\cite{Koniger2008,McCain1998,Dutta2015,Dutta2017,Jeong2002,Xiong2017,Gowacka2023}. 

% %As confinement is defined relative to the chloroplast size, larger chloroplasts are more confined, limiting their ability to efficiently pack and re-arrange within plant cells.
% Further investigation of the cell morphology revealed that the plant has only flat cuboid cells with varying length while their height and width remain approximately constant within the tissue. This suggests that leaf cells grow only in one direction rather than in all directions, as expected from development in monocot plants with simple leaf architecture, where growth is restricted to a zone close to the leaf base~\cite{Mathur2004,Nelissen2016}. We showed that this feature allows cells to remain in a region of optimal morphology during growth and development. \\
%  the chloroplast size and number
%  We then analyzed chloroplast size and number, which scale with cell size, as found in other plants\cite{Honda1971,Ellis1985,Pyke1992}. More intriguingly, we found that chloroplasts are approximately Gaussian distributed,
% Our study provides evidence that plants such as \textit{Elodea densa} possess a size-control mechanism for its epidermal leaf cells that finds an optimal solution for confinement geometries for disk packing.
% This intriguing interplay of cell development and organelle size and number selection displays a fertile ground for future investigations.\\
While the cell sizes and shapes align well with optimal packing solutions, the experimentally observed densities consistently fall below the shape-dependent maximal packing fraction by up to $10\%$. Several factors can explain this discrepancy: 1) cells are not perfectly cuboid, and chloroplasts are not perfectly disk-shaped; 2) chloroplasts are embedded within a cellular matrix of other organelles, cytoplasm, and especially cytoskeletal filaments~\cite{Wada2018,Kadota2009}, which impose an inter-chloroplast spacing, suggesting that the effective radius of chloroplasts might be slightly underestimated; and importantly 3) our previous study~\cite{schramma2023chloroplasts} suggested that chloroplasts are close to, but not deep into a glassy phase, allowing for space for re-arrangements, especially to facilitate efficient transitions between the two packing configurations. 
Notably, the packing fractions we found are rather close to a liquid-hexatic transition region ~\cite{Marcus1996,Bernard2011,Liu2021} which also shows reduced hexatic order (Fig.~\ref{fig:Fig3}e) as compared to the dense packing from simulations.
% % Our simulations random close packing of a mono-disperse distributed disks in strong confinement. We see non-monotonic deviations from the hyperbolic law in strong confinement, which were previously found in bi-disperse mixtures~\cite{Desmond2009}.
% % TBD
% We note that our protocol of generating packings \textit{in silico} differs largely from the dynamical chloroplast organization inside plant cells, which might have an impact on the achievable chloroplast packing structures. Despite this potential discrepancy between experiments and simulation, we are positive that the represented theoretical trends will be largely unaffected.\\

Overall, our study highlights the importance of packing problems in confinement in biological systems, which might be the key to understanding the collective light-controlled chloroplast re-arrangement within plant cells, a physiologically relevant process for light adaptation.
Our findings suggest that plant cells evolve and develop geometries that balance the motion and re-arrangement of organelles inside the cell with the structure demand of the tissue. This raises the intriguing question: how are cellular and developmental processes shaped by packing constraints across scales? 
While it is known that intracellular processes can be modulated via macromolecular crowding~\cite{VanDenBerg2017,Holt2023}, we expect that not only organelle size~\cite{Marshall2020} but also the packing configuration of organelles, condensates, and vesicles may play crucial roles for cell physiology, homeostasis and mechanics~\cite{Chang2023,Chen2024}.\\

To further broaden the scope of our work, we note that we mainly investigated the two chloroplast configurations within the plant cells. However, the transition and coordination of chloroplast motion offer another intriguing direction of research: chloroplasts can individually sense and move in response to light~\cite{Wada2018} and yet re-arrange collectively by coordinated movements similar to flocks, form three-dimensional aggregates and spread on the cell walls. Besides its biological relevance, this rich phenomenology displays fertile ground for future studies, especially from a perspective of phase transitions of confined active matter systems.

\section*{\textit{Materials and Methods}}

% \begin{figure}
% 	\centering
% 	\includegraphics[width=1.\textwidth]{Figures/SI_Fig_Statistics.pdf}
% 	\caption{\textbf{Geometrical analysis of cell size and packing density}. Multichannel imaging with (a) brightfield channel and (b) autofluorescence of chlorophyll showing disk-like chloroplasts. Star-dist segmentation of chloroplasts (c) and hand-segmented cell shapes (d) with indicated long and short axes. }
% 	\label{fig:CP_Packing_Elongation}
% \end{figure}

\subsection*{Imaging and image processing}
We measured the chloroplast density in 3 different leaves of \textit{Elodea densa} plants. For this purpose, we detached healthy leaves from the stem and imaged the bottom layer (abaxial layer) of the tissue. For imaging a Nikon TI2 microscope, a Prime BSI Express sCMOS camera and brightfield illumination using red-color bandpass filter  ($600\,\mathrm{nm}$, $FWHM=40\,\mathrm{nm}$) was used (Fig.~\ref{fig:SI_Fig3}a). Additionally chlorophyll autofluorescence imaging was performed using a CY5 filter cube (excitation: $604 - 644\,\mathrm{nm}$, emission: $672-712\,\mathrm{nm}$) (Fig.~\ref{fig:SI_Fig3}b).\\
To account for the curvature of the leaf tissue, we acquire z-stacks of $10-20\,\mathrm{\mu m}$.
Z-stacks were compressed into a single plane by extended depth-of-field-stacking using a Sobel filter approach (kernel width $=10\,\mathrm{px}$) to account for the curvature of the underlying tissue. Chloroplasts were segmented using StarDist~\cite{Weigert2020} (Fig.~\ref{fig:SI_Fig3}c).
Cells are segmented by hand from brightfield images (Fig.~\ref{fig:SI_Fig3}d). Segmented chloroplasts are assigned to their cells and filtered by size with an equivalent diameter $d=2\sqrt{A/\pi}$ in a range of $2-15\,\mathrm{\mu m}$; note that chloroplasts are expected to be around $4-6\,\mathrm{\mu m}$ in diameter. Additionally, we require a solidity (area divided by convex hull area) of above $0.7$ to ensure mostly convex particles, excluding mis-detections. 
Packing fractions were calculated by summing chloroplast areas in different cells: $\phi = \sum_{i=1}^N A_i /\mathcal{A}$. The p-atic order parameter (\ref{eq:hex}) was calculated from chloroplast positions via a distance matrix $D_{ij}$ and a cutoff depending on chloroplast radii $r_i$: $d_{ij}=2(r_i+r_j)+r_a$ where $r_a=0.4$ represents an additional zone of $20\%$ of the radius around every chloroplast. \\
We calculate the Voronoi tesselation of chloroplasts within each cell and measure the Voronoi volumes $\mathcal{V}$. We compare the measured histograms with the maximum entropy distribution $P(\mathcal{V}) = \frac{k^k}{\Gamma(k)}\frac{(\mathcal{V}-\mathcal{V}_c)^{k-1}}{(\langle\mathcal{V}\rangle-\mathcal{V}_c)^k} \exp\left(-k\frac{\mathcal{V}-\mathcal{V}_c}{\langle\mathcal{V}\rangle-\mathcal{V}_c}\right)$, where $\mathcal{V}_c \approx \min(\mathcal{V})$ and $k=\frac{\langle\mathcal{V} \rangle-\mathcal{V}_c}{ \mathrm{var}(\mathcal{V})}$.\\
Cell shape parameters such as aspect ratio, length scales $L_x$ and $L_y$, area $\mathcal{A}$ and perimeter $\mathcal{C}$ are calculated from the mask images.

\subsection*{Confocal imaging and image processing}
We mount an \textit{Elodea} leaf on a microscope slide with a spacer, remove the water from the aquarium culture, and immerse it in a mixture of one drop (approx $100\,\mathrm{\mu l}$) Calcofluor White Stain (MERCK) and one drop of $10\,\%$ potassium hydroxide, subsequently we place a coverslide. Calcofluor is used to stain cellulose in the cell walls~\cite{Bidhendi2020}.
Confocal imaging is performed with a Leica SP8 in the Leeuwenhoek Centre for Advanced Microscopy, Amsterdam. A $405\,\mathrm{nm}$ diode laser is used for the excitation of Calcofluor White, and the emission band is set from $450-520\,\mathrm{nm}$. Chlorophyll autofluorescence is excited with a Helium Neon laser at the $633\,\mathrm{nm}$ line, and the emission band ranges from $640$ to $740\,\mathrm{nm}$. We acquire z-stacks of $0.36\,\mathrm{\mu m}$ step size with a x-y pixel size of $0.3\,\mathrm{\mu m}$ using a $60\times$ oil-immersion objective (NA=$1.4$)

For processing, the Calcofluor channel is first slightly blurred using a Gaussian filter ($\sigma=1\,\mathrm{px}$), then binarized using Li's method in FIJI~\cite{FIJI}, then inverted (i.e. cells interior is $1$ and cell walls $0$). Subsequent morphological opening using a cube of $2$ pixels and distance-transform watershed segmentation using Manhattan-metric and a 6-connectivity generated to a well-separated label map for cells. Subsequently, we rejected labels touching the upper or lower boundary of the field of view, enabling us to avoid segmentation of the upper cell layer and noisy background.
We measure the average height in the border of all cells within one chloroplast diameter $2\langle r \rangle\approx 4.25\,\mathrm{\mu m}$ from the side wall.
\subsection*{Disk packing algorithm}
We use a disk packing algorithm based on previous works~\cite{Xu2005,Clarke1987} and further modified and applied in~\cite{Meer2024,Desmond2009}. In brief, we first select a fixed number of particles N drawn from a size distribution $P(r)$, with $P(r)$ chosen to be a Gaussian with polydispersity $\delta = 13.6\%$ to match the experiments.  The $N$ particles are randomly placed in a confined system with initial size $L_x$ and $L_y$ such that the initial area fraction $\phi = 0.01$ and no particles are allowed to overlap.  We then loop through the particles in random order, trying to expand each particle’s size by a small amount $a$, if that expansion does not cause any overlap with other particles or with the walls of the system.  We additionally try to move each particle a small distance in a random direction, again only if that displacement does not cause an overlap.  When all particles have been successfully expanded by $a$, the system size is rescaled ($L_x' = L_x/a$, $L_y' = L_y/a$), particle sizes rescaled down by $a$, particle positions within the box likewise rescaled. If too many trials occur without every particle successfully expanding, then the particle sizes are reset to the last rescaled value, $a$ is decreased, and the trials resume.  This continues until $a - 1 = 10^{-5}$, at which point the simulation is concluded.  Throughout the simulation, the aspect ratio $L_x/L_y$ is kept fixed, so to explore the necessary conditions, we vary $N$ from $30$ to $130$ and $L_x/L_y$ from $1.0$ to $10.0$.  In total, we run $23479$ simulations, with at least $5$ repetitions of each condition and in many cases more.

\section*{Acknowledgements}
We extend our gratitude to Karen Villari for her contributions to obtaining cell shape images.
We also thank Ronald Breedijk for his assistance with confocal microscopy at the Leeuwenhoek Centre for Advanced Microscopy, Amsterdam.
Additionally, we thank Max Bi, Kartik Chhajed, Marko Popović, Isabelle Eisenmann, Yuri Z. Sinzato, and Jared Popowski for their insightful discussions.
This material is based on work supported by the National Science Foundation under Grant No. CBET-2306371 (E.R.W.). M.J. acknowledges support from the ERC grant no.~"2023-StG-101117025, FluMAB." This publication is part of the Vidi project \emph{Living Levers} with file number 21239, financed by the Dutch Research Council (NWO). 

\section*{Author contribution}
Conceptualization N.S.;
Data curation N.S.;
Formal analysis N.S. and E.R.W.;
Funding acquisition: M.J.;
Investigation: N.S., E.R.W and M.J.;
Methodology \& Software: N.S. and E.R.W.;
Project administration: M.J.;
Supervision: M.J.;
Visualization: N.S. and M.J.;
Writing – original draft: N.S. and M.J.;
Writing – review \& editing: N.S., E.R.W. and M.J.
\printbibliography

\newpage
\section*{Appendix}
\setcounter{figure}{0}
\renewcommand\thefigure{S\arabic{figure}}   

% We analyzed the 3D confocal images of plant tissues to extract the height profile (Fig.~\ref{fig:SI_Fig1}). The resulting height maps feature peaks or ridges within the center of the cell, which coincide with the cell walls of the upper cell layer. The height around the side walls of the cell averages around $L_z\approx2.34\pm 1$. We used strong light to induce chloroplast rearrangement toward side walls (Fig.~\ref{fig:SI_Fig2}). The chloroplasts pack predominantly at side walls, but some clusters within the cell center are visible. These clusters typically align with upper cell walls or the position of chloroplasts in the upper cell layer. \\
% The cell length and width were quantified using brigthfield microscopy, combined with fluorescence microscopy to detect chloroplasts (Fig.~\ref{fig:SI_Fig3}). 

\begin{figure}[h!]
	\centering
	\includegraphics[width=1\textwidth]{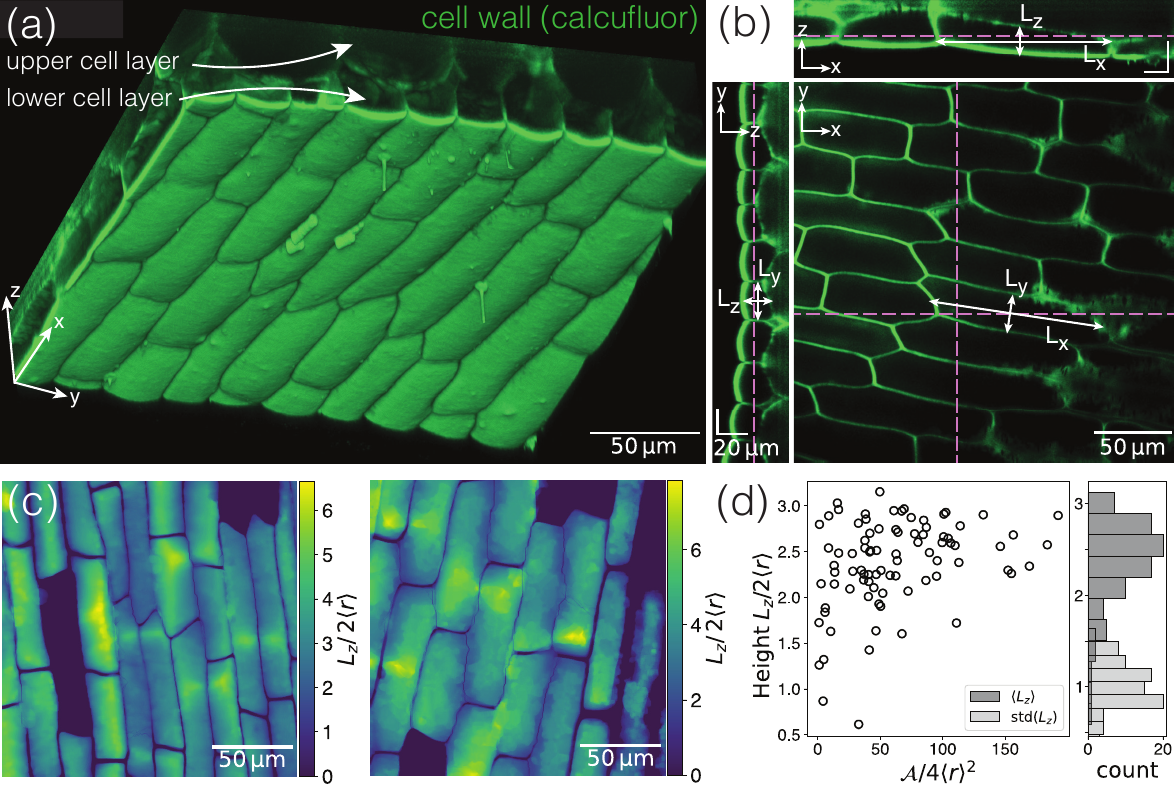}
	\caption{\textbf{3D image of cell walls} (a) 3D view of upper and lower cell walls using calcufluor (b) Orthogonal slices (along dashed lines). Definitions of cell length $L_x$, width $L_y$, and height $L_z$. (c) Height profile of various cells. Height peaks correspond to cell boundaries from the upper cell layer. (d) The height of cells lies around $L_z=2.35\pm1$ chloroplast diameters.}
	\label{fig:SI_Fig1}
\end{figure}
\begin{figure}[h!]
	\centering
	\includegraphics[width=1\textwidth]{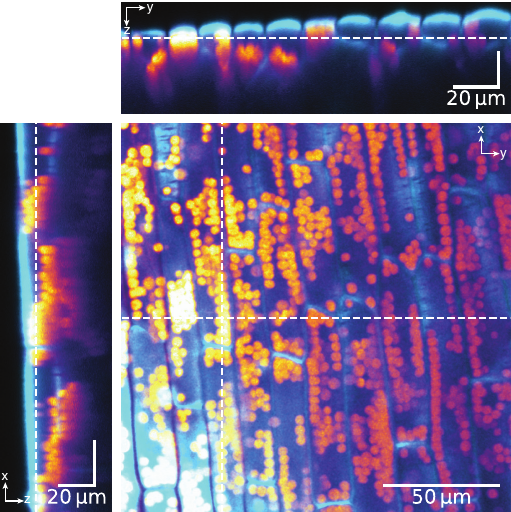}
	\caption{\textbf{Bright light avoidance configuration of chloroplasts.} Orthogonal view of cell walls (cyan) and chloroplasts (orange) along dotted white lines, respectively.
    $xy$-plane represents the maximum intensity picture of chloroplasts. Chloroplasts cluster on the side walls. Blob-like chloroplast clusters are visible. These co-localize with wider profiles along cell walls (yz view) of the upper cell layer or chloroplast aggregates in the upper layer (xz view).}
	\label{fig:SI_Fig2}
\end{figure}

\begin{figure}[h!]
	\centering
	\includegraphics[width=1\textwidth]{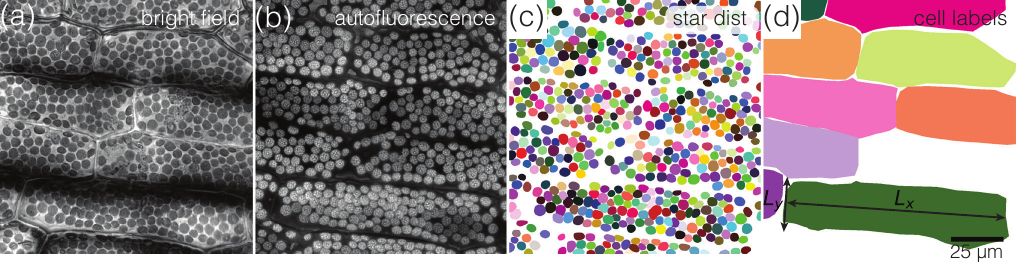}
	\caption{\textbf{Image analysis} (a) brightfield channel to detect cell walls (b) widefield fluorescence channel of chlorophyll. (c) Segmentation using StarDist\cite{Weigert2020}. (d) Hand-segmented cells with definitions of $L_x$ and $L_y$.}
	\label{fig:SI_Fig3}
\end{figure}
\begin{figure}[h!]
	\centering
	\includegraphics[width=1\textwidth]{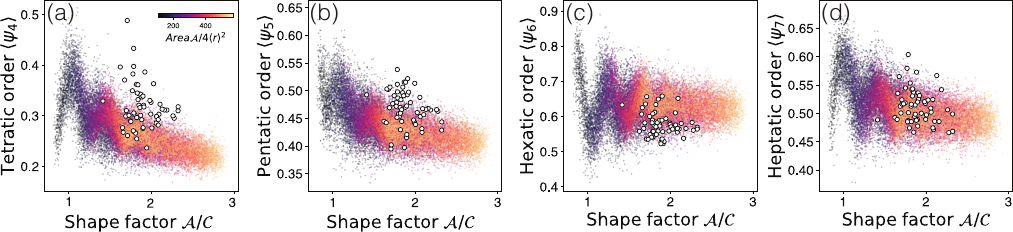}
	\caption{\textbf{p-atic order parameters}: (a)-(d) tetratic ($p=4$), pentatitc ($p=5$), hexatic ($p=6$) and heptatic $p=7$ order. Open symbols: experimental data. Color represents cthe onfinement area.}
	\label{fig:SI_Fig4}
\end{figure}
\begin{figure}[h!]
	\centering
	\includegraphics[width=1\textwidth]{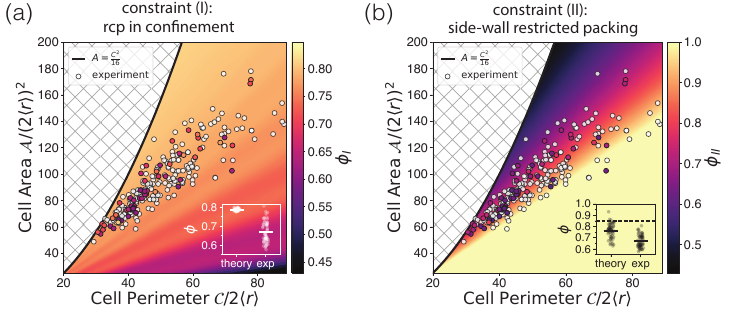}
	\caption{\textbf{Constraints (I) and (II)} (a) Random close packing in confinement corresponding to equation \eqref{eq:packing_geometry}. (b) Area-sidewall mismatch constraint, corresponding to equation \eqref{eq:packing_sidewalls}. Colored points correspond to experimental data, with colors representing the packing density respectively. Note that colormap limits at $\phi_{II}=1$, which is total coverage and not achievable by any disk packing. Insets: comparison of theoretical versus experimentally measured packing fraction $\phi$, suggesting that data is limited by both constraints.}
	\label{fig:SI_Fig5}
\end{figure}

\end{document}